\documentclass[aps,prstab,reprint,groupedaddress]{revtex4-2}

\usepackage{amsmath}
\usepackage{tikz}
\usepackage{siunitx}

\graphicspath{{./figures/}}

\begin{document}

\title{Tracking with wakefields in dielectric laser acceleration grating structures}

\author{Thilo Egenolf}
\email[]{egenolf@temf.tu-darmstadt.de}

\affiliation{Institut f\"ur Teilchenbeschleunigung und Elektromagnetische Felder (TEMF), Technische Universit\"at Darmstadt, Schlossgartenstrasse 8, D-64289 Darmstadt, Germany}

\author{Uwe Niedermayer}
\affiliation{Institut f\"ur Teilchenbeschleunigung und Elektromagnetische Felder (TEMF), Technische Universit\"at Darmstadt, Schlossgartenstrasse 8, D-64289 Darmstadt, Germany}

\author{Oliver Boine-Frankenheim}
\altaffiliation{Also at GSI Helmholtzzentrum f\"ur Schwerionenforschung GmbH, Planckstrasse 1, D-64291 Darmstadt, Germany}
\affiliation{Institut f\"ur Teilchenbeschleunigung und Elektromagnetische Felder (TEMF), Technische Universit\"at Darmstadt, Schlossgartenstrasse 8, D-64289 Darmstadt, Germany}

\date{November 8, 2019}

\begin{abstract}
Due to the tiny apertures of dielectric laser acceleration grating structures within the range of the optical wavelength, wakefields limit the bunch charge for relativistic electrons to a few femtocoulomb. In this paper, we present a wakefield upgrade of our six-dimensional tracking scheme DLAtrack6D in order to analyze these limitations. Simulations with CST Studio Suite provide the wake functions to calculate the kicks within each tracking step. Scaling laws and the dependency of the wake on geometrical changes are calculated. 
The tracking with wakefields is applied to beam and structure parameters following recently performed and planned experiments. We compare the results to analytical models and identify intensity limits due to the transverse beam breakup and strong head-tail instability. Furthermore, we reconstruct phase advance spectrograms and use them to analyze possible stabilization mechanisms.
\end{abstract}


\maketitle

\section{Introduction}
Dielectric laser acceleration (DLA) structures accelerate electrons in the optical near-fields of periodic gratings~\cite{Shimoda1962ProposalMaser,Lohmann1962ElectronWaves}. Powered by ultrafast lasers, they are a promising concept for compact accelerators due to tenfold higher gradients as compared to conventional RF accelerators~\cite{Peralta2013DemonstrationMicrostructure}. 
Recently, accelerating gradients of \SI{850}{\mega\electronvolt\per\metre} were demonstrated at UCLA~\cite{Cesar2018High-fieldAccelerator} using \SI{45}{\femto\second} laser pulses. Reviews of DLA theory and experiments can be found in~\cite{England2014DielectricAccelerators,Mcneur2016Laser-drivenAccelerator,Niedermayer2018ChallengesAcceleration}.
\begin{figure*}[htp]
\begin{minipage}[b]{0.45\textwidth}
\includegraphics[width=\textwidth]{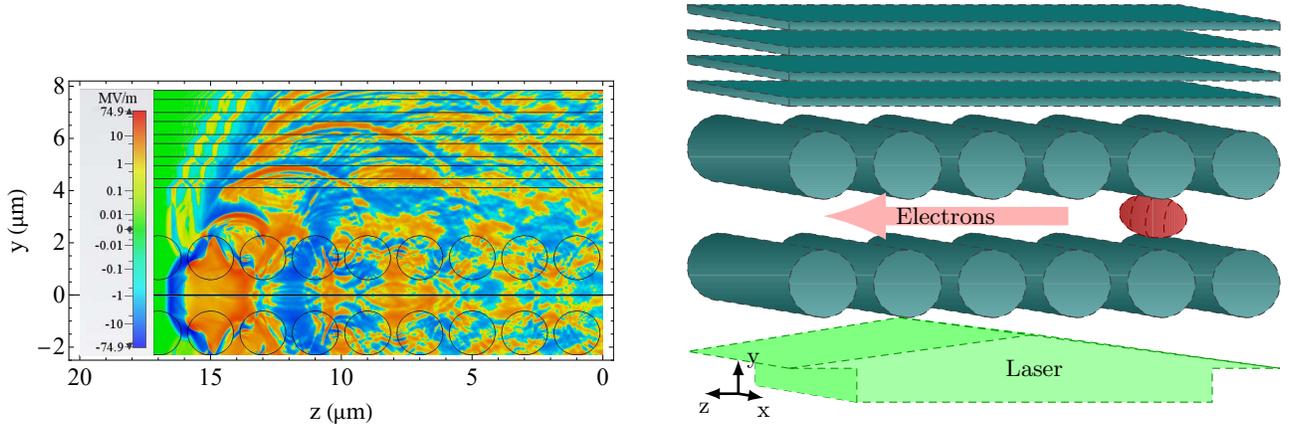}
\end{minipage}
\hspace{0.04\textwidth}
\begin{minipage}[b]{0.45\textwidth}
\begin{tikzpicture}[scale=0.45]
\foreach\y in {-1,...,2}
    {\fill[blue!50!green] (-1,6+0.75*\y) -- (13.5,6+0.75*\y) -- (13.5,6.25+0.75*\y) -- (10.5,6.75+0.75*\y) -- (-4,6.75+0.75*\y) -- (-4,6.5+0.75*\y) -- cycle;
    \fill[color=gray!90!,shading=axis,opacity=0.25] (-1,6+0.75*\y) rectangle (13.5,6.25+0.75*\y);
    \fill[gray!30!black,shading=axis,opacity=0.25] (-1,6+0.75*\y) -- (-1,6.25+0.75*\y) -- (-4,6.75+0.75*\y) -- (-4,6.5+0.75*\y) -- cycle;
    \fill[gray!50!black,opacity=0.25] (-1,6.25+0.75*\y) -- (13.5,6.25+0.75*\y) -- (10.5,6.75+0.75*\y) -- (-4,6.75+0.75*\y) -- cycle;
    \draw[densely dashed,gray!50!black] (-1,6+0.75*\y) rectangle (13.5,6.25+0.75*\y);
    \draw[densely dashed,gray!50!black] (-1,6+0.75*\y) -- (-4,6.5+0.75*\y) -- (-4,6.75+0.75*\y) -- (-1,6.25+0.75*\y);
    \draw[densely dashed,gray!50!black] (-4,6.75+0.75*\y) -- (10.5,6.75+0.75*\y) -- (13.5,6.25+0.75*\y);}

\def\y{0};
\foreach \x in {5,...,0}
    {\fill[blue!50!green](2.5*\x,\y+1) -- (2.5*\x-3,\y+1.5) arc (90:270:1) -- (2.5*\x,\y-1) arc (270:450:1);
    \fill[top color=gray!50!black,bottom color=gray!50!black,middle color=gray!50,shading=axis,opacity=0.25] (2.5*\x,\y+1) -- (2.5*\x-3,\y+1.5) arc (90:270:1) -- (2.5*\x,\y-1) arc (270:90:1);
    \fill[gray!90!,shading=axis,opacity=0.25] (2.5*\x,\y) circle (1);
    \draw[densely dashed,gray!50!black] (2.5*\x,\y) circle (1);
    \draw[densely dashed,gray!50!black] (2.5*\x,\y+1) -- (2.5*\x-3,\y+1.5)  arc (90:270:1) -- (2.5*\x,\y-1);}

\fill[red,rotate around={-9.46:(9.75,2)}] (9.75,2) circle (1 and 0.625);
\fill[left color=gray!90!, right color=gray!2!,rotate around={-9.46:(9.75,2)},opacity=0.25] (9.75,2) circle (1 and 0.625);
\draw[densely dashed,red!50!black,rotate around={-9.46:(9.75,2)}] (9.75,2) circle (1 and 0.625);
\draw[densely dashed,red!50!black,rotate around={-9.46:(9.75,2)}] (9.75,2.625) arc (90:270:0.2 and 0.625);
\draw[densely dashed,red!50!black,rotate around={-9.46:(9.75,2)}] (9.25,2.541) arc (90:270:0.15 and 0.541);
\draw[densely dashed,red!50!black,rotate around={-9.46:(9.75,2)}] (10.25,2.541) arc (90:270:0.15 and 0.541);

\def\y{3.5};
\foreach \x in {5,...,0}
    {\fill[blue!50!green](2.5*\x,\y+1) -- (2.5*\x-3,\y+1.5) arc (90:270:1) -- (2.5*\x,\y-1) arc (270:450:1);
    \fill[top color=gray!50!black,bottom color=gray!50!black,middle color=gray!50,shading=axis,opacity=0.25] (2.5*\x,\y+1) -- (2.5*\x-3,\y+1.5) arc (90:270:1) -- (2.5*\x,\y-1) arc (270:90:1);
    \fill[gray!90!,shading=axis,opacity=0.25] (2.5*\x,\y) circle (1);
    \draw[densely dashed,gray!50!black] (2.5*\x,\y) circle (1);
    \draw[densely dashed,gray!50!black] (2.5*\x,\y+1) -- (2.5*\x-3,\y+1.5)  arc (90:270:1) -- (2.5*\x,\y-1);}

\draw[densely dashed,green!60!black] (1,-3.5) -- (-2,-3) -- (-2,-2);
\fill[green!50!white] (-1,-2.5) -- (-4,-2) -- (2.25,-1) -- (6.25,-1.5) -- cycle;
\fill[green!50!white] (2.25,-1) -- (13.5,-2.5) -- (6.25,-1.5) -- cycle;
\fill[green!50!white] (1,-3.5) -- (-2,-3) -- (-2,-2) -- (1,-2.5) -- cycle;
\fill[green!35!white] (-1,-2.5) -- (6.25,-1.5) -- (13.5,-2.5) -- (11.5,-2.5) -- (11.5,-3.5) -- (1,-3.5) -- (1,-2.5) -- cycle;
\draw[densely dashed,green!60!black] (-1,-2.5) -- (6.25,-1.5) -- (13.5,-2.5) -- (11.5,-2.5) -- (11.5,-3.5) -- (1,-3.5) -- (1,-2.5) -- cycle;
\draw[densely dashed,green!60!black] (-1,-2.5) -- (-4,-2) -- (2.25,-1) -- (6.25,-1.5);
\draw[densely dashed,green!60!black] (2.25,-1) -- (13.5,-2.5);
\node at (6.25,-2.5) {Laser};

\draw[->, >=latex, red!30!white, line width=7pt]   (7.25,2) to node[black]{Electrons} (0,2);

\draw[->, >=latex,black,line width=1pt] (-2.5,-4.25+1) -- (-3.5,-4.25+1);
\node[below,black] at (-3.5,-4.25+1) {z};
\draw[->, >=latex,black,line width=1pt] (-2.5,-4.25+1) -- (-2.5,-3.25+1);
\node[right,black] at (-2.5,-3.25+1) {y};
\draw[->, >=latex,black,line width=1pt] (-2.5,-4.25+1) -- (-1.75,-4.375+1);
\node[below,black] at (-1.75,-4.375+1) {x};
\end{tikzpicture}
\end{minipage}
\caption{\label{fig:cstFieldLong}Left: Longitudinal electric field of a \SI{50}{\nano\metre} bunch (at $z=\SI{16.565}{\micro\metre}$) with \SI{1}{\femto\coulomb} bunch charge (left) propagating through a dielectric dual pillar structure with \SI{2}{\micro\metre} period length (right)}
\end{figure*}
The optimization of the beam dynamics in DLA is dominated by the question of how to fit a finite emittance beam into the sub-micrometer apertures of the grating structures over a length of several thousand periods. A scheme to confine a beam in the longitudinal as well as in one transverse plane by alternating the synchronous phase (APF-scheme) was proposed recently~\cite{Niedermayer2018Alternating-PhaseAcceleration}. In this work, we described the single electron dynamics of an accelerator, which is fully scalable in length and energy, however intensity effects were not included, yet. These intensity effects play a major role in relativistic DLA experiments, where conventional photo-injectors and booster-linacs are used.

Coherent DLA acceleration requires bunch lengths significantly shorter than the optical wavelength. Together with the small apertures in DLAs, this leads to a strong short-range wakefield compared to the laser fields. As an example, Fig.~\ref{fig:cstFieldLong} shows the wakefield in a dielectric dual pillar structure. Previous work on the wake effects in DLAs has concentrated on longitudinal effects such as beam loading. This has been described by simplified analytical models of the structures, namely an azimuthal symmetrical and longitudinally flat structure and a pointlike bunch distributions ~\cite{Schachter2004EnergyCollider,Siemann2004EnergyAccelerators,Hanuka2014BraggOptimization,Hanuka2018OptimizedBunch,Hanuka2018OptimizedBunchb}. The analysis of metallic periodically corrugated structures like flat grating-dechirpers~\cite{Bane2016AnalyticalDechirper} has shown that the geometrical parameters influence the wakefields of particles close to the structure. Similar outcomes are expected for dielectric structures, however, additionally to the Smith-Purcell effect also the Cherenkov effect is present. 

Due to the finite emittance of available sources, the small apertures of a DLA grating are almost completely filled by particles. In order to analyze intensity effects in arbitrary DLA grating structures, we expand our tracking scheme DLAtrack6D~\cite{Niedermayer2017BeamScheme} by kicks due to the charge-distribution dependent wake in this paper.

DLAtrack6D is based on calculating the longitudinal kicks from the spatial Fourier harmonics of the laser field and deriving the transverse kicks by means of the Panofsy-Wenzel theorem~\cite{Panofsky1956SomeFields}. Subsequently, the kicks are applied to each particle of a bunch and the tracking is performed by the symplectic Euler (or Leap Frog) method. In this paper, we add the wake kicks in the a similar way as the external laser kicks.
The paper is organized as follows: The calculation of the wake-kicks is summarized in Sec.~\ref{sec:WakeKicks}, which is followed by a description of the exemplarily simulated grating structures and scaling laws in Sec.~\ref{sec:StructureAndScaling}. Sec.~\ref{sec:SimulationResults} shows tracking results for different energy ranges and links them to simplified analytical models of transverse instabilities. Stabilization mechanisms and intensity limits are also given in this section. The paper concludes with a summary and an outlook in Sec.~\ref{sec:Conclusion}.

\section{Wake kicks}
\label{sec:WakeKicks}
\begin{figure}[h]
    \centering
    \begin{tikzpicture}[scale=1.2]
        \draw[fill=gray!20] (-1.5,-1) rectangle (-1,1);
        \node [rotate=90] (pillar) at (-1.25,0) {pillar};
        \draw[fill=gray!20] (1,-1) rectangle (1.5,1);
        \node [rotate=90] (pillar) at (1.25,0) {pillar};

        \draw [->] (-2.5,0) -- (-2.5,0.5) node [left] {x};
        \draw [->] (-2.5,0) -- (-2,0) node [below] {y};
        \draw (-2.5,0) circle (0.1) node [below left] {z};
        \draw (-2.5-0.0707,0.0707) -- (-2.5+0.0707,-0.0707);
        \draw (-2.5-0.0707,-0.0707) -- (-2.5+0.0707,0.0707);
            
        \fill [blue] (0.75,0) circle (0.05);
        \fill [blue] (0.5,0) circle (0.05);
        \fill [blue] (0.25,0) circle (0.05);
        \fill [blue] (0,0) circle (0.05);
        \fill [blue] (-0.25,0) circle (0.05);
        \fill [blue] (-0.5,0) circle (0.05);
        \fill [blue] (-0.75,0) circle (0.05);
        \fill [blue] (0.75,-0.25) circle (0.05);
        \fill [blue] (0.5,-0.25) circle (0.05);
        \fill [blue] (0.25,-0.25) circle (0.05);
        \fill [blue] (0,-0.25) circle (0.05);
        \fill [blue] (-0.25,-0.25) circle (0.05);
        \fill [blue] (-0.5,-0.25) circle (0.05);
        \fill [blue] (-0.75,-0.25) circle (0.05);
        \fill [blue] (0.75,-0.5) circle (0.05);
        \fill [blue] (0.5,-0.5) circle (0.05);
        \fill [blue] (0.25,-0.5) circle (0.05);
        \fill [blue] (0,-0.5) circle (0.05);
        \fill [blue] (-0.25,-0.5) circle (0.05);
        \fill [blue] (-0.5,-0.5) circle (0.05);
        \fill [blue] (-0.75,-0.5) circle (0.05);
        \fill [blue] (0.75,-0.75) circle (0.05);
        \fill [blue] (0.5,-0.75) circle (0.05);
        \fill [blue] (0.25,-0.75) circle (0.05);
        \fill [blue] (0,-0.75) circle (0.05);
        \fill [blue] (-0.25,-0.75) circle (0.05);
        \fill [blue] (-0.5,-0.75) circle (0.05);
        \fill [blue] (-0.75,-0.75) circle (0.05);
        \fill [blue] (0.75,0.25) circle (0.05);
        \fill [blue] (0.5,0.25) circle (0.05);
        \fill [blue] (0.25,0.25) circle (0.05);
        \fill [blue] (0,0.25) circle (0.05);
        \fill [blue] (-0.25,0.25) circle (0.05);
        \fill [blue] (-0.5,0.25) circle (0.05);
        \fill [blue] (-0.75,0.25) circle (0.05);
        \fill [blue] (0.75,0.5) circle (0.05);
        \fill [blue] (0.5,0.5) circle (0.05);
        \fill [blue] (0.25,0.5) circle (0.05);
        \fill [blue] (0,0.5) circle (0.05);
        \fill [blue] (-0.25,0.5) circle (0.05);
        \fill [blue] (-0.5,0.5) circle (0.05);
        \fill [blue] (-0.75,0.5) circle (0.05);
        \fill [blue] (0.75,0.75) circle (0.05);
        \fill [blue] (0.5,0.75) circle (0.05);
        \fill [blue] (0.25,0.75) circle (0.05);
        \fill [blue] (0,0.75) circle (0.05);
        \fill [blue] (-0.25,0.75) circle (0.05);
        \fill [blue] (-0.5,0.75) circle (0.05);
        \fill [blue] (-0.75,0.75) circle (0.05);
            
        \fill [red] (0.05,0) arc (0:180:0.05) -- cycle;
        \fill [red] (0.3,0) arc (0:180:0.05) -- cycle;
        \fill [red] (0.55,0) arc (0:180:0.05) -- cycle;
        \fill [red] (0.8,0) arc (0:180:0.05) -- cycle;
        \fill [red] (-0.2,0) arc (0:180:0.05) -- cycle;
        \fill [red] (-0.45,0) arc (0:180:0.05) -- cycle;
        \fill [red] (-0.7,0) arc (0:180:0.05) -- cycle;
\end{tikzpicture}
    \caption{\label{fig:cstSimulationGrid}Integration paths (blue dots) and beam paths (red dots) in the CST simulation. For each beam path the simulation is performed and the resulting wakefields are recorded in every integration path.}
\end{figure}
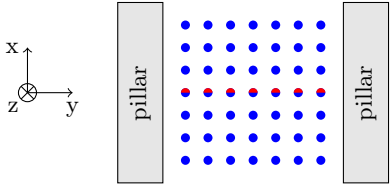

The calculation of wake kicks requires solving Maxwell's equations in a given structure. Starting from a periodic grating structure, which is translation invariant in $x$-direction, we simulate the wake potential by the time-domain wakefield solver in CST Studio Suite~\cite{CST2018CST2018} (cf. Fig.~\ref{fig:cstFieldLong}). The wakefield-integration paths are arranged on a rectangular grid in the gap region and the beam path is varied along the locations of the integration path in $y$-direction (cf. Fig.~\ref{fig:cstSimulationGrid}). The simulation is performed for each position of the source beam. Due to the translation symmetry of the structure in $x$-direction, it is sufficient to have all beam paths at only one $x$-position.
If the length of the Gaussian excitation signal is much shorter than the real bunch length, the simulation wake result is a good approximation of Green's function $\vec{w}\left(x-\tilde{x},y,\tilde{y},\tilde{s}\right)$, where $\tilde{x}$ and $\tilde{y}$ are the coordinates of the beam path, $x$ and $y$ are the coordinates of the integral path, and $\tilde{s}$ is the longitudinal coordinate with $\tilde{s}=0$ at the position of the charge. The convolution with an arbitrary bunch distribution on an arbitrary transverse position in the gap results in the wake potential
\begin{align}
\begin{split}
    \vec{W}\left(x,y,s\right)=\int_{-\infty}^\infty\int_{-\infty}^\infty\int_{-\infty}^\infty\left[\vec{w}\left(x-\tilde{x},y,\tilde{y},\tilde{s}\right)\right.\\
    \left.\times\lambda\left(\tilde{x},\tilde{y},s-\tilde{s}\right)\right]d\tilde{x}d\tilde{y}d\tilde{s},
    \label{eq:wakeFunction}
\end{split}
\end{align}
where the bunch distribution is normalized as $\int_V\lambda\left(x,y,s\right)dV=1$. For simplicity, we assume an uncorrelated distribution $\lambda\left(\tilde{x},\tilde{y},s-\tilde{s}\right)=\lambda_x\left(\tilde{x}\right)\lambda_y\left(\tilde{y}\right)\lambda_s\left(s-\tilde{s}\right)$. This splits the integral in Eq.~\ref{eq:wakeFunction} in three parts, which can be evaluated independently. The wake potential is the result of a convolution over the longitudinal coordinate
\begin{equation}
    \vec{W}\left(x,y,s\right)=\int_{-\infty}^\infty\vec{W}_{xy}\left(x,y,\tilde{s}\right)\lambda_s\left(s-\tilde{s}\right)d\tilde{s},
\end{equation}
where the integrals over the transverse coordinates are
\begin{equation}
    \vec{W}_{xy}\left(x,y,\tilde{s}\right)=\int_{-\infty}^\infty\vec{W}_{y}\left(x-\tilde{x},y,\tilde{s}\right)\lambda_x\left(\tilde{x}\right)d\tilde{x}
\end{equation}
and
\begin{equation}
    \vec{W}_{y}\left(x-\tilde{x},y,\tilde{s}\right)=\int_{-\infty}^\infty\vec{w}\left(x-\tilde{x},y,\tilde{y},\tilde{s}\right)\lambda_y\left(\tilde{y}\right)d\tilde{y}.
\end{equation}
We will now discretize the above integrals, i.e. approximate them as sums. The integral over the transverse coordinate $\tilde{y}$ is, for example, approximated as
\begin{align}
\begin{split}
    \vec{W}_y\left(x-\tilde{x},y,\tilde{s}\right)\approx&\Delta\tilde{y}\sum_{n_{\tilde{y}}}\left[\vec{w}\left(x-\tilde{x},y,n_{\tilde{y}} \Delta\tilde{y},\tilde{s}\right)\right.\\
    &\left.\times\lambda_y\left(n_{\tilde{y}} \Delta\tilde{y}\right)\right].
\end{split}
\end{align}
The sum can be interpreted as a tensor product with all coordinates replaced by representative indices
\begin{subequations}
\begin{align}
    \vec{W}_y\left(x-\tilde{x},y,\tilde{s}\right)&\approx\vec{W}_{n_x,n_{\tilde{x}},n_y,n_{\tilde{s}}}^y\\
    &=\Delta\tilde{y}\sum_{n_{\tilde{y}}}\vec{w}_{n_x,n_{\tilde{x}},n_y,n_{\tilde{y}},n_{\tilde{s}}}\lambda^y_{n_{\tilde{y}}}.
    \label{eq:tensor}
\end{align}
\end{subequations}
The discrete representation of Eq.~\ref{eq:wakeFunction} follows as
\begin{widetext}
\begin{equation}
    \vec{W}\left(n_x,n_y,n_s\right) =\Delta\tilde{s}\Delta\tilde{x}\Delta\tilde{y}\sum_{n_{\tilde{s}}}\sum_{n_{\tilde{x}}}\sum_{n_{\tilde{y}}}\vec{w}_{n_x,n_{\tilde{x}},n_y,n_{\tilde{y}},n_{\tilde{s}}}\lambda^y_{n_{\tilde{y}}}\lambda^x_{n_{\tilde{x}}}\lambda^s_{n_{\tilde{s}},n_s},
\end{equation}
\end{widetext}
which can be implemented in a very efficient manner.

\begin{figure*}[htp]
\includegraphics[width=0.95\textwidth]{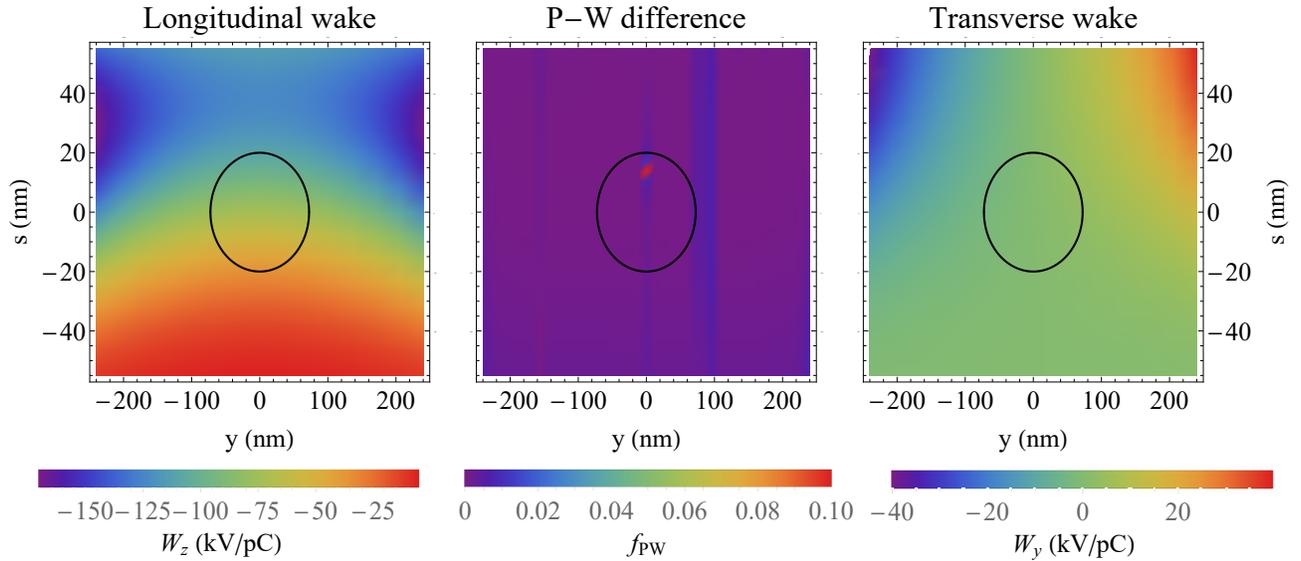}
\caption{\label{fig:wakePW}Longitudinal and transverse wake potential per grating period in the left and right plot, respectively. The plot in the center shows the normalized difference between the left end the right hand side of the Panofsky-Wenzel theorem (Eq.~\ref{eq:wakePWf}) calculated for the simulated data of a Gaussian bunch distribution with standard deviation given by the black ellipse.\\}
 \end{figure*}

The calculation of the wake kicks in each grating period starts with particle binning to determine the bunch distribution. Three tensor products as shown in Eq.~\ref{eq:tensor} yield the position dependent discrete wake potential for the bunch passing through one grating period. Evaluating the wake potential at the particle coordinates $(x_p,y_p,s_p)$, the wake kicks for each particle can be calculated as
\begin{subequations}
\begin{align}
    \centering
    \Delta x'\left(x_p,y_p,s_p\right)&=\frac{q\, q_\mathrm{bunch}}{p_{z0}\beta_\mathrm{ref}c_0}W_x\left(x_p,y_p,s_p\right),\\
    \Delta y'\left(x_p,y_p,s_p\right)&=\frac{q\, q_\mathrm{bunch}}{p_{z0}\beta_\mathrm{ref}c_0}W_y\left(x_p,y_p,s_p\right),\\
    \Delta\delta\left(x_p,y_p,s_p\right)&=\frac{q\,q_\mathrm{bunch}}{E_{0,\mathrm{ref}}}W_s\left(x_p,y_p,s_p\right)
\end{align}
\label{eq:kicks}
\end{subequations}
with the particle charge $q$, the bunch charge $q_\mathrm{bunch}$, the speed of light $c_0$, the reference energy $E_{0,\mathrm{ref}}$, the reference velocity $\beta_\mathrm{ref}$, and the reference momentum $p_{z0}$. The changes in transverse and longitudinal momenta can be combined by the Panofsky-Wenzel theorem~\cite{Panofsky1956SomeFields}
\begin{equation}
\partial_s\Delta\vec{p}_\bot=-\nabla_\bot\Delta p_s,
\label{eq:pw}
\end{equation}
which is applicable under periodic boundary conditions~\cite{Niedermayer2017BeamScheme}. Inserting the kicks (Eq.~\ref{eq:kicks}) with $\Delta x'=\Delta p_x / p_{z0}$, $\Delta y'=\Delta p_y / p_{z0}$ and $\Delta\delta=\Delta p_s\beta c / E_{0,\mathrm{ref}}$, Eq.~\ref{eq:pw} becomes
\begin{equation}
   \partial_s\vec{W}_\bot=-\nabla_\bot W_s,  
   \label{eq:wakePW}
\end{equation}
i.e. the same property as for the laser kicks also holds for the wake kicks.
Equation~\ref{eq:wakePW} is a generally known result in wake field research (see e.g.~\cite{Palumbo1994WakeImpedance}); here we use it as validation of Eqs~\ref{eq:kicks} and also of the numerical simulations and post-processing steps. 
The relative error in fulfilling Eq.~\ref{eq:pw} is written as 
\begin{equation}
    f_\textrm{PW}\left(x,y,s\right)=\frac{\left|\frac{\partial W_s}{\partial y}\left(x,y,s\right)+\frac{\partial W_y}{\partial s}\left(x,y,s\right)\right|}{\left|\frac{\partial W_y}{\partial s}\left(x,y,s\right)\right|+\left|\frac{\partial W_s}{\partial y}\left(x,y,s\right)\right|}
    \label{eq:wakePWf}
\end{equation}
with $f_\textrm{PW}\left(x,y,s\right)\ll1$, which is plotted in Fig.~\ref{fig:wakePW}. 
The plot in the center shows that the numerical difference in the Panofsky-Wenzel theorem is well below 1\% except for one spot where both derivatives of the wake are zero and thus artifacts of the finite differences blow up the relative error.

The wake kicks (Eqs~\ref{eq:kicks}) are added to the kicks by the external laser field (cf. Eq. 23 in~\cite{Niedermayer2017BeamScheme}). The prior implementation of DLAtrack6D runs in \textsc{Matlab}~\cite{MathWorks2016Matlab}. For the presented extension, DLAtrack6D was migrated to \textsc{Python}~\cite{PythonSoftwareFoundation2018Python3.7}. All convolutions are implemented as matrix multiplications, which makes the calculation fast and efficient, such that it can run on an ordinary PC.

\section{Structures and Scaling}
\label{sec:StructureAndScaling}

Equipped with the simulation tools to calculate the wakefields and the resulting kicks we investigate an example of a relativistic DLA structure as shown in Fig.~\ref{fig:cstFieldLong}. It consists of two rows of pillars made of fused silica ($\varepsilon_r=2.13$) with a Bragg mirror on one side to symmetrize the external accelerating fields~\cite{Yousefi2019DielectricReflector}. The vacuum gap as channel for the electrons between the two pillar rows has a width of \SI{1.2}{\micro\metre}. The period length equals the laser wavelength of \SI{2}{\micro\metre}, i.e. the structure is matched to $\beta=1$. The dimensions of the pillars are optimized in terms of acceleration gradient and can be found in Appendix~\ref{sec:appDimensions}.
 
If we change the geometric parameters of a structure, we have to repeat the expensive wakefield simulations for each parameter change. This effort can be significantly reduced by using scaling laws. If all dimensions of the grating structure are modified by a scaling factor $\alpha$, i.e. $\tilde{s}=\alpha s$, the Green's function wake per period scales as $\alpha^{-1}$~\cite{Bane1987WakefieldCollider}, i.e. $\tilde{w}\left(\tilde{s}\right)=\alpha^{-1}w\left(s\right)$. For a longitudinal bunch distribution, which is also scaled as $\tilde{\lambda}\left(\tilde{s}\right)=\alpha^{-1}\lambda\left(s\right)$, the wake potential scales the same as the Green's function wake. This follows from the convolution
\begin{subequations}
\begin{align}
    \tilde{W}\left(\tilde{s}\right)&=\int_{-\infty}^\infty\tilde{\lambda}\left(\tilde{s}\right)\tilde{w}\left(\tilde{s}\right)\\
    &=\alpha^{-1}\int_{-\infty}^\infty\lambda\left(s\right)w\left(s\right)\\
    &=\alpha^{-1}W\left(s\right).
\end{align}
\end{subequations}

Scaling the period length of a grating structure and the laser wavelength for $\beta\approx1$ is almost equivalent to modifying all dimensions since a change in period length causes a change in acceptable bunch length for coherent acceleration and also a change in the decay length of the evanescent fields in the channel. 
We will use this to scale the wakefields of DLA experiments performed at $\lambda_0=$ \SI{800}{\nano\metre} to the same structure at $\lambda_0=$ \SI{2}{\micro\metre}, i.e. $\alpha^{-1}=2.5$.

\begin{figure}[htp]
\includegraphics[width=0.45\textwidth]{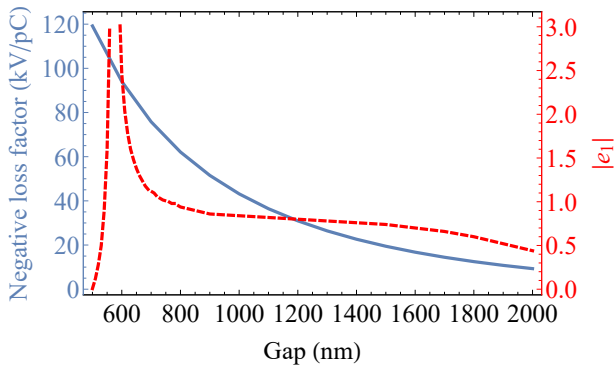}
\caption{\label{fig:gapVariation}Loss factor indicating the strength of wake effects as function of the vacuum gap compared to the absolute value of the first Fourier coefficient indicating the accelerating gradient in the center of a structure for a given external field strength.}
\end{figure}

The loss factor is defined as
\begin{equation}
    k_\textrm{loss}=\int\limits_{-\infty}^\infty\int\limits_{-\infty}^\infty\int\limits_{-\infty}^\infty W_s\left(x,y,s\right)\lambda\left(x,y,s\right)dxdyds.
\end{equation}
An increase of the gap width lowers the wakefield and thus the loss factor significantly. However, the acceleration gradient expressed by the first Fourier coefficient $e_1$ of the external field in the center of the gap also decreases with the gap width. Figure~\ref{fig:gapVariation} shows the loss factor indicating the strength of the wake effects in comparison to $|e_1|$ for the grating described previously as function of the gap width. 

The plateau in the first Fourier coefficient represents a robust optimum for the vacuum gap in the range of the previously chosen value of \SI{1.2}{\micro\metre}. Values in the resonant range, e.g. \SI{0.6}{\micro\metre}, are not useful as they are not robust and also impede the use of short laser pulses due to the limited bandwidth.

\begin{figure}[htp]
\includegraphics[width=0.45\textwidth]{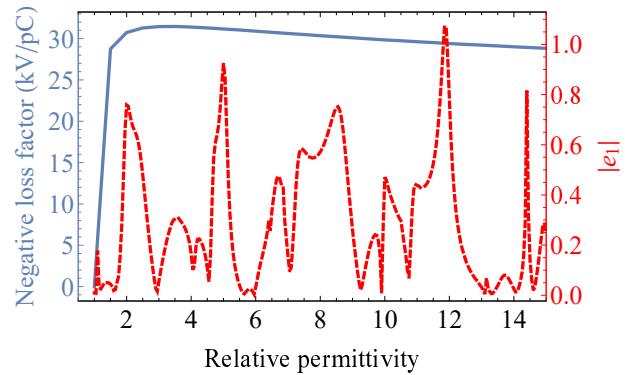}
\caption{\label{fig:epsVariation}Loss factor indicating the strength of wake effects as function of the relative permittivity of the dielectric compared to the absolute value of the first Fourier coefficient indicating the accelerating gradient in the center of a structure for a given external field strength.}
\end{figure}

In contrast to the vacuum gap, changing the relative permittivity only slightly affects the wake strength due to its broadband excitation. However, since the laser field has a small bandwidth approximated as single frequency excitation, $|e_1|$ is very sensitive to that. The loss factor and the first Fourier coefficient as functions of the relative permittivity are plotted in Fig.~\ref{fig:epsVariation}. This indicates that wake considerations do not have to be taken into account for the choice of material but only in the optimization of the structure.

Recent experiments use different dual-layer rectangular grating structures~\cite{Niedermayer2017DesigningChip,Cesar2018High-fieldAccelerator}. If we optimize the teeth of a fused-silica dual-layer rectangular grating with the same vacuum gap as the analyzed dual pillar structure in terms of acceleration gradient, we reach a gradient \SI{40}{\percent} lower than for the dual pillar structure. The dimensions of the optimized structure are given in Appendix~\ref{sec:appDimensions}. Figure~\ref{fig:comparisonPillarRec} compares the short range longitudinal wake of both grating structures. The curve shapes are almost the same. The wake of the dual pillar structure is, however, slightly smaller due to the rounded material transition at the gap boundary.

\begin{figure}[htp]
\begin{tikzpicture}
\definecolor{mathematicablue}{rgb}{0.3684,0.5068,0.7098};
\definecolor{mathematicayellow}{rgb}{0.8807,0.611,0.1421};

\node[inner sep=0pt] (plot) at (0,0)
    {\includegraphics[width=0.45\textwidth]{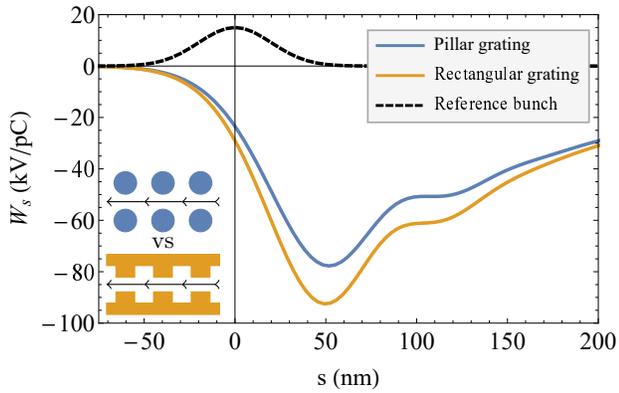}};

\def\offsetx{-1.5};
\def\offsety{0};

\foreach \x in {0,...,2}
    {\draw[mathematicablue, fill=mathematicablue] (\x/2-1+\offsetx,-0.3+\offsety) circle (0.15);
    \draw[mathematicablue, fill=mathematicablue] (\x/2-1+\offsetx,0.2+\offsety) circle (0.15);}
\draw[<-<] (-0.25+\offsetx,-0.05+\offsety) -- (0.25+\offsetx,-0.05+\offsety);
\draw[<-] (-0.75+\offsetx,-0.05+\offsety) -- (-0.25+\offsetx,-0.05+\offsety);
\draw[<-] (-1.25+\offsetx,-0.05+\offsety) -- (-0.75+\offsetx,-0.05+\offsety);

\node[text=black] at (-0.5+\offsetx,-0.6+\offsety) {vs};

\draw[mathematicayellow, fill=mathematicayellow] (-1.25+\offsetx,-0.75+\offsety) rectangle (0.25+\offsetx,-0.9+\offsety);
\draw[<-<] (-0.25+\offsetx,-1.15+\offsety) -- (0.25+\offsetx,-1.15+\offsety);
\draw[<-] (-0.75+\offsetx,-1.15+\offsety) -- (-0.25+\offsetx,-1.15+\offsety);
\draw[<-] (-1.25+\offsetx,-1.15+\offsety) -- (-0.75+\offsetx,-1.15+\offsety);
\draw[mathematicayellow, fill=mathematicayellow] (-1.25+\offsetx,-1.4+\offsety) rectangle (0.25+\offsetx,-1.55+\offsety);
\foreach \x in {0,...,2}
    {\draw[mathematicayellow, fill=mathematicayellow] (\x/2-1-0.125+\offsetx,-0.9+\offsety) rectangle (\x/2-1+0.125+\offsetx,-1.05+\offsety);
    \draw[mathematicayellow, fill=mathematicayellow] (\x/2-1-0.125+\offsetx,-1.25+\offsety) rectangle (\x/2-1+0.125+\offsetx,-1.4+\offsety);}

\end{tikzpicture}
\caption{\label{fig:comparisonPillarRec}Longitudinal wake per grating period of a Gaussian bunch distribution passing a dual pillar structure (blue) and a dual-layer rectangular grating structure (yellow) on-axis. The reference bunch distribution is shown in dashed black.}
\end{figure}

\begin{figure}[htp]
\begin{tikzpicture}
\definecolor{mathematicablue}{rgb}{0.3684,0.5068,0.7098};
\definecolor{mathematicayellow}{rgb}{0.8807,0.611,0.1421};

\node[inner sep=0pt] (plot) at (0,0)
    {\includegraphics[width=0.45\textwidth]{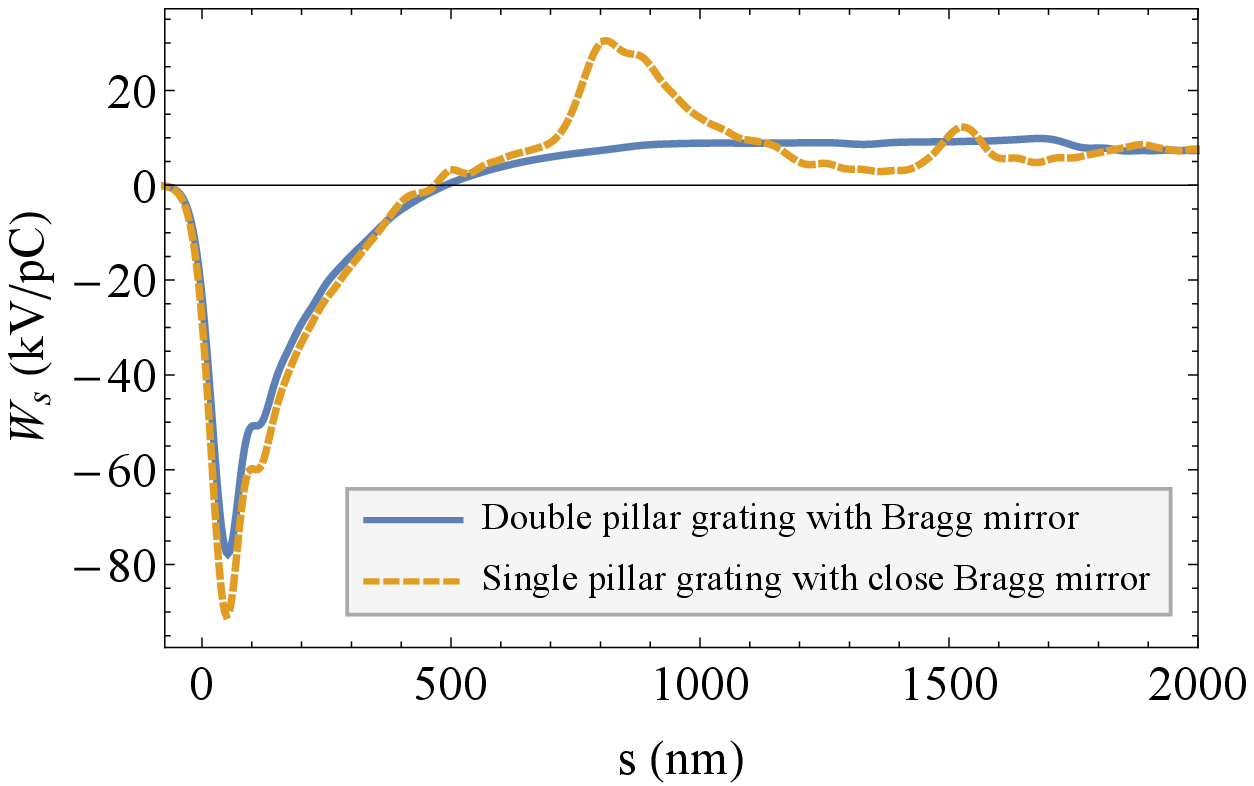}};

\def\offsetx{1.25};

\node[text=black] at (0.5+\offsetx,0) {vs};

\foreach \x in {0,...,2}
    {\draw[mathematicablue, fill=mathematicablue] (\x/2-1+\offsetx,-0.3) circle (0.15);
    \draw[mathematicablue, fill=mathematicablue] (\x/2-1+\offsetx,0.2) circle (0.15);}
\draw[mathematicablue, fill=mathematicablue] (-1.25+\offsetx,0.5) rectangle (0.25+\offsetx,0.6);
\draw[mathematicablue, fill=mathematicablue] (-1.25+\offsetx,0.7) rectangle (0.25+\offsetx,0.8);
\draw[mathematicablue, fill=mathematicablue] (-1.25+\offsetx,0.9) rectangle (0.25+\offsetx,1);
\draw[<-<] (-0.25+\offsetx,-0.05) -- (0.25+\offsetx,-0.05);
\draw[<-] (-0.75+\offsetx,-0.05) -- (-0.25+\offsetx,-0.05);
\draw[<-] (-1.25+\offsetx,-0.05) -- (-0.75+\offsetx,-0.05);

\foreach \x in {0,...,2}
    {\draw[mathematicayellow, fill=mathematicayellow] (\x/2-1+2+\offsetx,-0.3) circle (0.15);}
\draw[mathematicayellow, fill=mathematicayellow] (-1.25+2+\offsetx,0.05) rectangle (0.25+2+\offsetx,0.15);
\draw[mathematicayellow, fill=mathematicayellow] (-1.25+2+\offsetx,0.25) rectangle (0.25+2+\offsetx,0.35);
\draw[mathematicayellow, fill=mathematicayellow] (-1.25+2+\offsetx,0.45) rectangle (0.25+2+\offsetx,0.55);
\draw[<-<] (-0.25+2+\offsetx,-0.05) -- (0.25+2+\offsetx,-0.05);
\draw[<-] (-0.75+2+\offsetx,-0.05) -- (-0.25+2+\offsetx,-0.05);
\draw[<-] (-1.25+2+\offsetx,-0.05) -- (-0.75+2+\offsetx,-0.05);
\end{tikzpicture}
\caption{\label{fig:comparisonBraggDistanceZ}Longitudinal wake per grating period of a Gaussian bunch distribution passing a dual pillar structure with a Bragg mirror (blue) and a single-row pillar structure, where the second row is replaced by a Bragg mirror (yellow) on-axis.}
\end{figure}

\begin{figure}[htp]
\includegraphics[width=0.45\textwidth]{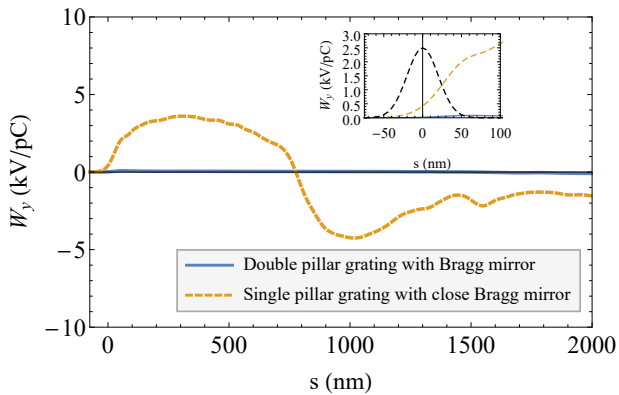}
\caption{\label{fig:comparisonBraggDistanceY}Transverse wake per grating period of a Gaussian bunch distribution passing a dual pillar structure with a Bragg mirror(blue) and a single-row pillar structure, where the second row is replaced by a Bragg mirror (yellow) on-axis. The inset shows an enlargement of the short-range wake and the reference bunch distribution in black.}
\end{figure}

Furthermore, the Bragg mirror breaks the symmetry of the grating structure. If the distance of the Bragg mirror is much larger than the vacuum gap, it does not affect the short-range wake and can thus have only multi-bunch effects. If the Bragg mirror is located near to the vacuum gap or one pillar row is even replaced by the Bragg mirror, it affects also the short-range wake (cf. Fig.~\ref{fig:comparisonBraggDistanceZ}). Breaking the symmetry generates especially a non-vanishing transverse wake deflecting an on-axis bunch (cf. Fig.~\ref{fig:comparisonBraggDistanceY}).


\section{Transverse Wake Effects}
 \label{sec:SimulationResults}

\subsection{Previous estimations}
In~\cite{Egenolf2019IntensityStructures} we have calculated the wakefields and the beam loading limit for the dual pillar structure described previously. The tracking results of a transversely small bunch verify these results in the range of a few femtocoulombs. However we have also shown, that the wake of bunches which fill the whole aperture of the structure depends strongly on the transverse position of each source particle. That has created the demand to add 3D wakefields to DLAtrack6D in order to analyze transverse instabilities.

\subsection{Beam breakup}
\begin{figure}[htp]
    \begin{tikzpicture}
    \node[inner sep=0pt] (plot) at (0,0)
        {\includegraphics[width=0.45\textwidth]{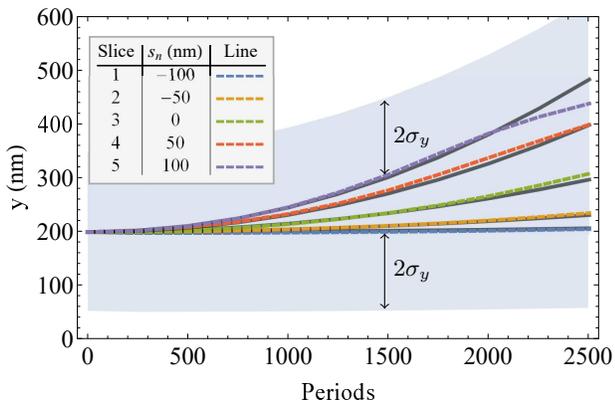}};
    \draw[<->] (1,0.41) -- (1,1.39) node [midway,right]{$2\sigma_y$};
    \draw[<->] (1,-0.38) -- (1,-1.38) node [midway,right]{$2\sigma_y$};
    \end{tikzpicture}
    \caption{\label{fig:sliceOffset}Comparison of slice center of mass for a tracked bunch divided into five slices (dashed lines) with the analytical description (black solid lines). The gray area shows the $2\sigma$ distance to the minimum and maximum curve. The aperture is at $y=\SI{600}{\nano\metre}$.}
\end{figure}

\begin{figure}[htp]
    \includegraphics[width=0.45\textwidth]{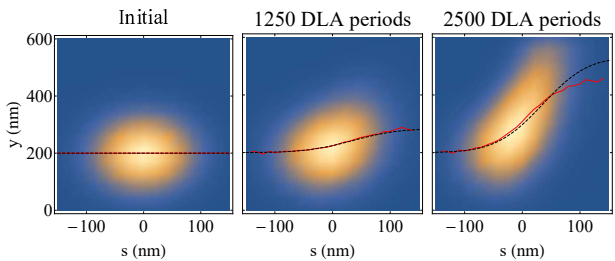}
    \caption{\label{fig:3gevdistribution}Initial bunch distribution and bunch distribution after 1250 and 2500 DLA grating periods. The red curves show the means of transverse slices in comparison to the analytical estimates (black curves).}
\end{figure}

Caused by the potential at constant synchronous phase, the particles in a bunch undergo synchrotron motion. The synchrotron frequency is~\cite{Niedermayer2017DesigningChip}
\begin{equation}
    \Omega_s=\sqrt{\frac{-2\pi eG\sin{\left(\phi_s\right)}}{\lambda_0\beta^3\gamma^3m_e}},
\end{equation}
where $G$ is the accelerating peak gradient, $\phi_s$ the synchronous phase, $\lambda_0$ the laser wavelength, $\beta$ and $\gamma$ the relativistic factors and $m_e$ the electron rest mass. Consider parameters achievable at the DLA experiment at SwissFEL~\cite{Prat2017OutlineSwissFEL}, i.e. a \SI{3}{\giga\electronvolt} electron beam accelerated with a peak gradient of $G=$ \SI{1}{\giga\electronvolt\per\metre} and a synchronous phase $\phi_s=\SI{135}{\degree}$, the synchrotron frequency is $\Omega_s\approx\SI{43.9e6}{\per\second}$. The synchrotron period is thus $\lambda_s\approx\SI{42.87}{m}$. On the other hand, the DLA grating is outlined to be only \SI{1.5}{\milli\metre} long. Consequently, the synchrotron and betatron motion can be considered frozen within the DLA grating. However, transverse wakes can cause particle loss by defocusing or deflection. This can be described analytically by dividing the bunch distribution $\lambda\left(s\right)$ in $N$ slices with a center at $s_n$ and a width $\Delta s$. For a given initial offset $\hat{y}_0$ of the whole bunch, a deflecting force acts on each slice given by
\begin{equation}
y''_n=\frac{q_eq_\mathrm{bunch}}{p_{z0}\beta_\mathrm{ref}c_0\lambda_z}W_y\left(s_n,\hat{y}_0\right).
\end{equation}
The n-th slice has an offset of
\begin{equation}
    y_n\left(z\right)=\hat{y}_0+\frac{q_eq_\mathrm{bunch}}{2p_{z0}\beta_\mathrm{ref}c_0\lambda_z}W_y\left(s_n,\hat{y}_0\right)z^2
    \label{eq:sliceOffsets}
\end{equation}
with the longitudinal position z of the bunch. This is equal to Chao's two-particle model~\cite{Chao1993PhysicsAccelerators} in the limit $k_\beta z=2\pi z/\lambda_\beta=2\pi z/\lambda_s\ll1$. Figure~\ref{fig:sliceOffset} shows the tracking results of a \SI{16}{fC} bunch injected with an offset of \SI{200}{nm} and the corresponding analytical solutions given by Eq.~\ref{eq:sliceOffsets} which fit quite well. For this, the bunch is divided into five slices and each is plotted separately.
As long as the bunch (pictured by the width of two standard deviations as gray area in the plot) is within the aperture, in this example up to about \SI{4.72}{\milli\metre} (2360 periods) grating length, the analytical description indicates negligible particle loss. This maximal interaction length is proportional to $\sqrt{p_{z0}\beta_\mathrm{ref}/q_\mathrm{bunch}}$. However, particle loss starts earlier as visible in the decreasing slope of the maximum tracked curve in Fig.~\ref{fig:sliceOffset}. This is caused by an increasing standard deviation of the bunch distribution due to defocusing of the macroparticle slices which is not included in the analytical description. Figure~\ref{fig:3gevdistribution} shows the initial and resulting bunch distributions after 1250 and 2500 DLA grating periods.
If the transverse width of a bunch is small compared to the aperture, the analytical description gives a good estimate for the transverse deflection of the bunch tail. However, if the bunch fills most of the aperture, tracking covers also the effect of the incoherent wake defocusing forces. 

To reduce the deflection and defocusing particle loss, on knob would be to avoid an injection offset as good as possible. However, this is technically difficult to ensure and helps only in symmetric structures. Generally, mitigation can only be obtained by reduced bunch charge, geometrical wake optimization, or simply by a more stiff beam at higher energy. 

\subsection{\label{sec:stabilization}Strong head-tail instability}
Considering a lower energy beam, for example a \SI{6.5}{\mega\electronvolt} beam accelerated with a gradient of \SI{0.56}{\giga\electronvolt\per\metre} corresponding to the experimental parameters at PEGASUS~\cite{Cesar2018EnhancedLaser}, the synchrotron period length is in the range of a few millimeters, i.e. in the same order of magnitude as the interaction length or below. Therefore, we have to take the longitudinal and transverse motion into account. Using the APF-scheme to confine the beam, the frequencies of longitudinal motion and transverse betatron motion are equal by construction~\cite{Niedermayer2018Alternating-PhaseAcceleration}. This distinguishes a DLA from conventional high energy linacs and also from synchrotrons, where the transverse betatron frequency is significantly larger than the longitudinal synchrotron frequency. For a validation of the tracking results including wake effects, we adapt the analytical description of the strong head-tail instability, in particular Chao's two-particle model for synchrotrons~\cite{Chao1993PhysicsAccelerators}, to these assumptions (see Appendix~\ref{sec:appBeamTransport}). In smooth approximation, the stability criterion given by the two-particle model is (cf. Eq.~\ref{eq:AppStabilitySmooth})
\begin{equation}
    \frac{q_eq_\mathrm{bunch}}{p_{z0}\beta_\mathrm{ref}c_0\lambda_z\hat{y}_0}W_y\left(\sigma_s,\hat{y}_0\right)\leq\frac{16\pi}{L_\beta^2}
    \label{eq:stabilitySmooth}
\end{equation}
with the betatron period length $L_\beta$ and assuming a wake which depends linearly on the transverse offset. The betatron period length is $L_\beta=2\pi L/\mu$, where L is the length of an APF FD-cell and 
\begin{equation}
    \mu=\arccos\left[\cos\left(\frac{k_\beta L}{2}\right)\cosh\left(\frac{k_\beta L}{2}\right)\right]
    \label{eq:phaseAdvance}
\end{equation}    
is the corresponding phase advance per APF cell~\cite{Niedermayer2018Alternating-PhaseAcceleration}. For the PEGASUS parameters, a FD-cell length of 1808 DLA periods (\SI{1.44}{\milli\metre}) with the structure in Fig.~\ref{fig:cstFieldLong} scaled to a laser wavelength of \SI{0.8}{\micro\metre} and a bunch length of \SI{20}{\nano\metre}, the maximal bunch charge is about \SI{0.3}{\femto\coulomb}. According to the APF design procedure in~\cite{Niedermayer2018Alternating-PhaseAcceleration}, the FD-cell length is chosen such that the maximal beta function (without wake) is minimized. Following Eq.~\ref{eq:stabilitySmooth}, a further minimization of the betatron period length would increase the maximal bunch charge. It would, however, increase the maximal beta function at the same time and thus decrease the initial acceptance. 

\begin{figure}[htp]
    \includegraphics[width=0.45\textwidth]{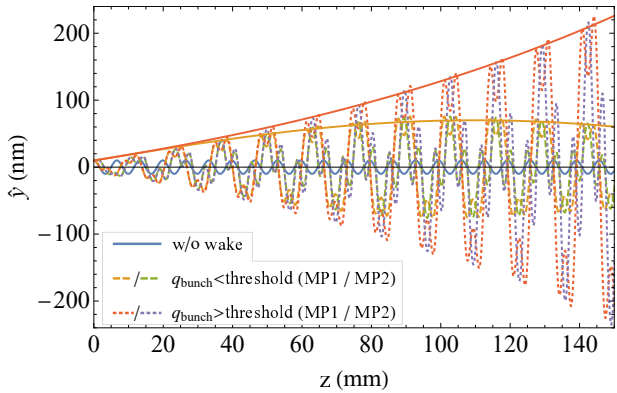}
    \caption{\label{fig:yOffsetStabilitySmooth}Two-particle model in smooth approximation. For a small initial offset, a transversely stable bunch has a sinusoidal envelope (yellow line) and a unstable bunch has an exponentially growing amplitude (red line). The y-offsets of both macroparticles (MP1 and MP2) are plotted respectively. The initial excitation without wake is shown for comparison (blue).}
\end{figure}

\begin{figure}[htp]
    \includegraphics[width=0.45\textwidth]{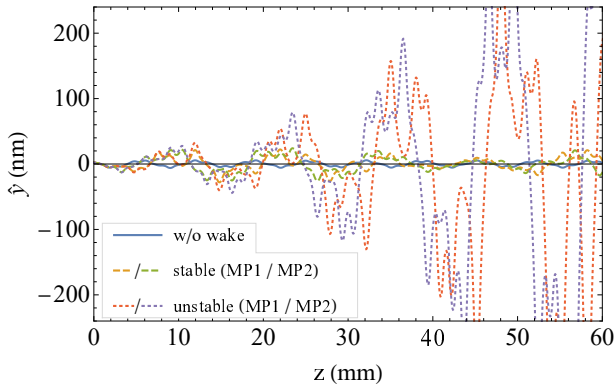}
    \caption{\label{fig:yOffsetStabilityAPF} Two-particle model using the APF transfer matrices.
    The y-offsets of both macroparticles (MP1 and MP2) are plotted respectively. The initial excitation without wake is shown for comparison (blue).}
\end{figure}

\begin{figure}[htp]
    \includegraphics[width=0.45\textwidth]{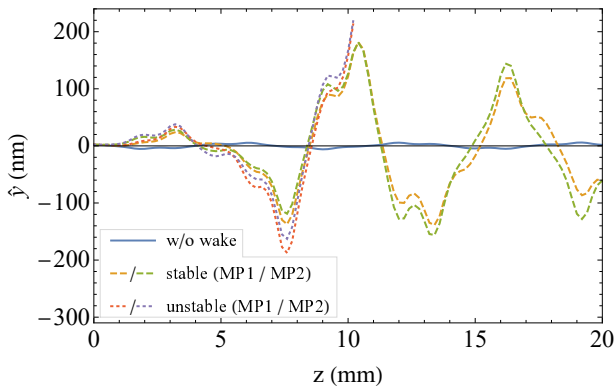}
    \caption{\label{fig:yOffsetStabilityTracking}Bunch distribution divided in two halves (MP1 and MP2) and tracked through a dual pillar APF grating. The betatron motion without wake is shown for comparison (blue).}
\end{figure}

Figure~\ref{fig:yOffsetStabilitySmooth} shows the $y$-offset of the two macroparticles along the  grating without wake and for a bunch charge below and above the threshold in Eq.~\ref{eq:stabilitySmooth} using the smooth approximation. The oscillation of a transversely stable bunch has a sinusoidal envelope whereas the oscillation amplitude of a unstable bunch grows exponentially.
In comparison, Fig.~\ref{fig:yOffsetStabilityAPF} shows the $y$-offset of the macroparticles calculated with the full APF transfer matrices given in Eqs.~\ref{eq:AppMatrixFW} and~\ref{eq:AppMatrixDW}.
The alternating sign of the focusing function reduces the stability threshold in this numerical example and for the analyzed interaction length roughly by half. The full tracking results shown in Fig.~\ref{fig:yOffsetStabilityTracking} confirm these semi-analytical estimates.
Thus, the two-particle model in smooth approximation can be seen as an upper limit for the estimation of the stability threshold. The more precise limit for a given APF grating with given interaction length is, however, smaller and can be estimated numerically using the presented tracking scheme.

\subsection{\label{sec:acceleration}Acceleration}
So far, we have not considered coherent acceleration of a bunch. A linear energy gain ${\gamma\left(z\right)=\gamma_0\left(1+G\cos{\phi_s}/\left(m_ec_0^2\right) z\right)}$ is usually provided for an ultrarelativistic accelerator design, i.e. peak acceleration gradient $G$ and synchronous phase $\phi_s$ are constant. The focusing strength decreases proportional to $\gamma^{-3/2}(z)$~\cite{Niedermayer2018Alternating-PhaseAcceleration} as compared to $\gamma^{-1/2}$ for a magnetic quadrupole lattice. To keep the phase advance per APF-cell constant, the length of the cell needs to be increased accordingly. This is equivalent to keeping the maximal beta function minimal, as was described in~\cite{Niedermayer2018Alternating-PhaseAcceleration}.

The corresponding equations of motion for the two-particle model are solved in Appendix~\ref{sec:app2} and the solutions are compared to Chao's two-particle model of a conventional linac. In Chao's model, the magnetic focusing strength is increased proportionally to the bunch energy to get a constant betatron wave number~\cite{Chao1993PhysicsAccelerators}. 
Increasing the focusing strength in an APF lattice is only possible at the price of increasing the laser field strength limited by material damage or playing with the synchronous phase. The adiabatic damping is being counteracted by the increase in the betafunction due to reduced focusing strength and thus a coherent offset of a bunch already increases without deflection by the wake (cf. Eq.~\ref{eq:AppSolutiony1}). Therefore, the adiabatic damping of an initial coherent offset in a conventional linac which leads to a prediction of an increased stability threshold~\cite{Chao1982CoherentBeam} does not help for DLA. Instead, the stability threshold is decreased compared to an APF transport channel.

 We confirmed that by tracking simulations, which are stable in the simulated interaction length in the case of transport and unstable for an accelerated bunch. For an acceleration from \SI{6.5}{\mega\electronvolt} up to \SI{17}{\mega\electronvolt}, the numerical simulations show a reduction of the threshold roughly by half as compared to an APF transport channel at constant energy.

\subsection{\label{sec:phaseAdvance}Analysis of the phase advance}
\begin{figure}[htp]
    \includegraphics[width=0.45\textwidth]{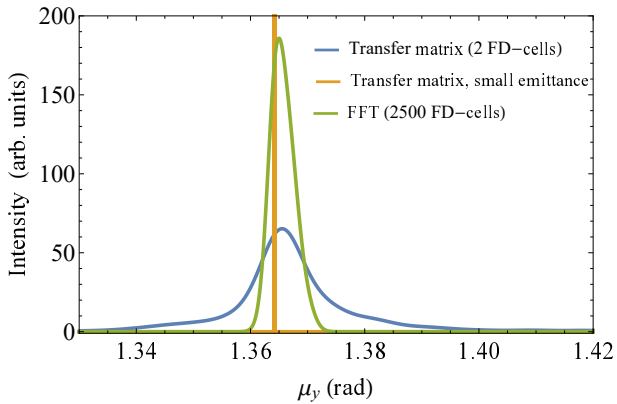}
    \caption{\label{fig:phaseAdvance}
    Transverse phase advance for a negligible emittance in complete agreement with the analytical $\mu_y=1.3642$ and a distribution with \SI{20}{\nano\metre} bunch length and \SI{100}{\nano\metre} transverse size. The latter is obtained by both FFT and the transfer matrix reconstruction method. The narrowing of the FFT curve originates from high amplitude particle losses over the long (2500 cells) transport distance.}
\end{figure}

The nonlinear spatial dependence of the fields in the dielectric grating leads to a phase advance depending on the particle's amplitude ('tune-spread'). 
The linear phase advance ('set-tune') is given analytically by Eq.~\ref{eq:phaseAdvance}. Numerically, the phase advance can be determined by reconstructing the transfer matrix of a FD-cell from the tracking result for each particle using the transfer matrix reconstruction algorithm presented in~\cite{Luccio2003EigenvaluesMatrix}. Figure~\ref{fig:phaseAdvance} shows the distributions of the transverse phase advance for a particle distribution with negligible emittance, which exactly confirms the analytical value $\mu_y=1.3642$, and a realistic emittance, where the distribution shows the expected broadening. Note that we calculate only two-dimensional transfer matrices neglecting coupling between planes. Calculating the phase advance by a Fast Fourier Transform (FFT) of particle tracking data along 2500 FD-cells yields the same mean value of the phase advance spectrum. However, tracking through such a long structure takes a lot of time and particle loss occurs. In particular, particles with large amplitudes have a large tune deviation and are likely to be lost, which leads to a narrowing of the spectrum. 

Physically, the spread is explained as follows: A larger longitudinal emittance leads to a weaker focusing and thus to a spread towards smaller phase advances. A larger transverse emittance, on the other hand, leads to a stronger focusing and thus to a spread towards larger phase advances.

\begin{figure}[htp]
    \includegraphics[width=0.45\textwidth]{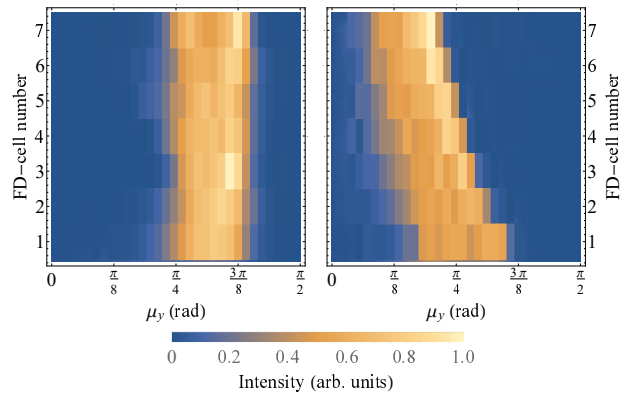}
    \caption{\label{fig:phaseAdvanceOverTime}Spectrograms of the transverse phase advance for a particle distribution accelerated in an APF structure. Increasing the FD-cell length along the structure in accordance to the increasing energy keeps the phase advance distribution almost constant (left). A constant FD-cell length, however, leads to a decreasing phase advance (right).}
\end{figure}

Furthermore, the calculation of the phase advance as trace of the transfer matrix allows a moving window approach to obtain a tune-spectrogram, in order to analyze the temporal variation of the phase advance e.g, in case of acceleration. As described in Sec.~\ref{sec:acceleration}, the length of a FD-cell varies over the structure length to keep the phase advance per FD-cell constant and thus the beam envelope bounded. The left plot in Fig.~\ref{fig:phaseAdvanceOverTime} shows the spectrum of the transverse phase advance for a accelerator design with increasing FD-cell length such that the designed linear phase advance on the acceleration ramp remains constant. The right plot shows the spectrum for a constant FD-cell length, where the phase advance decreases over time. The width of the distributions is composed of both the intrinsic width of the particle distribution and the numerical error that results from the calculation of the transfer matrix using two subsequent FD-cells, which are only approximately but not exactly identical.

\subsection{Damping mechanisms}
The upper bunch charge limit due to longitudinal wake effects of the analyzed structure is in the range of a few femtocoulombs~\cite{Egenolf2019IntensityStructures}, where the beam loading cancels the laser field with \SI{1}{\giga\electronvolt} acceleration gradient completely. As the stability criterion of the strong head-tail instability, however, is below a femtocoulomb for the same peak gradient, the bunch intensity is thus limited by the transverse effects. To raise the limits, a damping mechanism is required. 

BNS-damping~\cite{Balakin1983VLEPPDynamics} as routinely used in relativistic RF linacs is not applicable, as it requires a chirp in phase advance depending on the longitudinal position of the particles within the bunch. Due to the longitudinal motion in APF, the particle positions change continuously and a constant chirp in phase advance as function of longitudinal coordinate cannot be achieved.

\begin{figure}[htp]
    \includegraphics[width=0.45\textwidth]{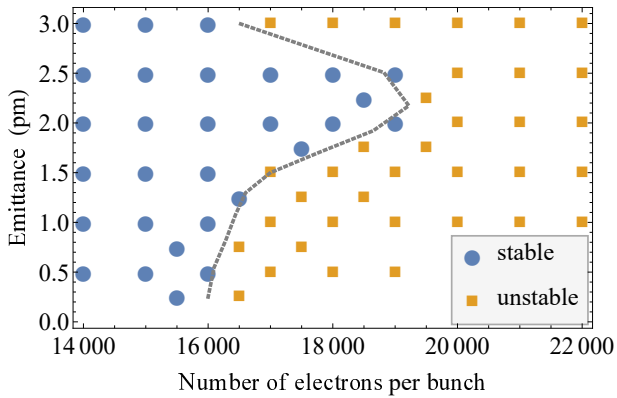}
    \caption{\label{fig:stabilityDiagram}Stability diagram of bunch distribution with varying transverse geometric emittance and bunch charge. The energy of the particles is \SI{6.5}{\mega\electronvolt}. Simulation settings with stable behaviour are colored yellow and settings with unstable behaviour are colored blue. The dotted line displays the interpolated limit between stable and unstable settings.}
\end{figure}

Another stabilizing mechanism is based on phase mixing~\cite{Hofmann2006LandauDamping}, i.e., the incoherent betatron frequencies spread could possibly stabilize the bunch against the strong head-tail instability. Thus, the bunch charge limit given by Eq.~\ref{eq:stabilitySmooth} is increased at larger emittances. In order to confirm that, tracking simulations through a \SI{5}{\centi\metre} long dual pillar APF grating with varying initial transverse emittances are performed. The matched Gaussian bunch distribution has an initial longitudinal length of \SI{20}{\nano\metre}, a constant width of \SI{400}{\nano\metre} in $x$-direction, and a particle energy of \SI{6.5}{\mega\electronvolt}. The FD-cell length is chosen so that the maximum of the Twiss beta parameter is minimal, i.e. $\hat{\beta}_\textrm{y,max}=\SI{0.69}{\milli\metre}$, and the assumed peak acceleration gradient is \SI{5}{\giga\electronvolt\per\metre}. The stable and unstable distributions as function of the transverse emittance and bunch charge are shown in Fig.~\ref{fig:stabilityDiagram}. In the limit of zero transverse emittance the limit of stability is already larger than described by the analytical stability criterion in Eq.~\ref{eq:stabilitySmooth} ($q_\textrm{bunch,max}\approx\SI{1.8}{\femto\coulomb}$), since stabilization is already provided by the finite longitudinal emittance. Moreover, the non-zero width in the $x$-direction weakens the wake. For increasing transverse emittance, the stability limit is shifted to higher bunch charges, which indicates a stabilization by phase mixing is really present here.
Due to the finite aperture of the grating, a further increase of the emittance leads to particle loss before the stabilization by phase mixing can be efficacious. Stable beam transport with larger charge and larger emittance is, however, possible with an increased aperture. This reduces on the other hand again the acceleration gradient, or requires more laser power, which is however limited by the material damage threshold. 

\section{Conclusion}
\label{sec:Conclusion}
We have successfully integrated wake kicks in our DLA particle tracker DLAtrack6D and presented for the first time tracking results with wakefields in nanophotonic electron acceleration structures. The wake functions themselves originate from electromagnetic simulations by the CST Particle Studio wakefield solver. When excited by a sufficiently short bunch, the obtained wake potential is a good approximation of the Green's function wake. We showed, that scaling laws can be applied to vary geometric parameters of the dielectric grating structures without complete recalculation of the wake. 

The tracking results showed, that transverse wake effects limit the bunch charge stronger than longitudinal effects. In the simulated example at \SI{6.5}{\mega\electronvolt}, the strong head-tail instability limits the bunch charge for stable beam transport to a few femtocoulomb for a reasonable peak gradient. At higher energies, \SI{3}{\giga\electronvolt} in the example, the synchrotron motion is too slow to have any stabilizing effect. The transverse wake deflects a non-centered bunch and 
the length before it hits the wall can be calculated upon the wake and the bunch charge. In both regimes (with or without longitudinal motion), the description of the transverse effects by analytical models, in particular by Chao's two-particle model, was confirmed by tracking simulations. 

We also calculated the spread in phase advance of a particle distribution moving through an APF DLA grating and numerically showed that phase mixing stabilizes the bunch in a DLA against transverse wake instabilities. Increasing the bunch charge further is only possible if the structure geometry is changed to make room for a larger emittance or if the focusing strength is increased. 

An increase of the vacuum gap would lower the wakefield but also the acceleration gradient. A more significant reduction of the wake would be the increase of the grating period length to the wavelength of CO lasers~\cite{Kimura2019COAccelerator}, that is \SI{10}{\micro\metre}, or to the  Terahertz range~\cite{Nanni2015Terahertz-drivenAcceleration}, where however the availability of efficient power sources represents the bottleneck.

In case of acceleration, the adiabatic damping of a coherent offset in a conventional RF linac with a magnetic focusing lattice increases the stability threshold of transverse instabilities. We showed that in a APF DLA adiabatic damping is counteracted by an increasing betafunction and a coherent offset increases. Thus, the stability threshold decreases for acceleration compared to a pure guiding structure. 

So far, we have not considered the particle motion in the $x$-direction which is assumed to be invariant in most DLA gratings. The transverse quadrupole wake leads to a focusing force in this direction depending on the width of the bunch. This has to be analyzed in more detail in the future, where also possible confinement methods in this direction have to be taken into account. 

\begin{acknowledgments}
This work is funded by the Gordon and Betty Moore Foundation (Grant No. GBMF4744) and the German Federal Ministry of Education and Research (Grant No. FKZ: 05K16RDB).
\end{acknowledgments}

\appendix
\section{Structure dimensions}
\label{sec:appDimensions}
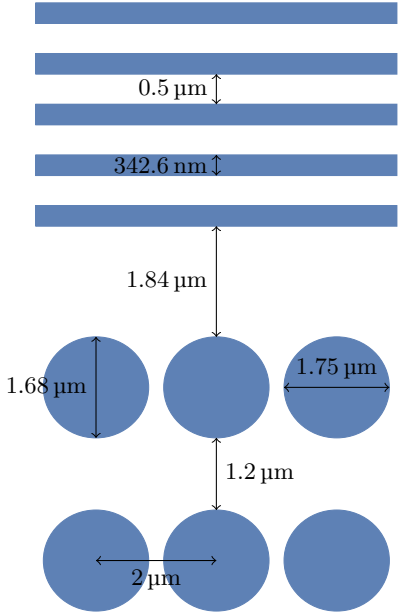
\begin{figure}[htp]
\begin{tikzpicture}[scale=0.8]
\definecolor{mathematicablue}{rgb}{0.3684,0.5068,0.7098};
\definecolor{mathematicayellow}{rgb}{0.8807,0.611,0.1421};

\def\pillarradiusy{0.84};
\def\pillarradiusz{0.875};
\def\periodlength{2};
\def\gap{1.2};
\def\braggdistance{1.84};
\def\lbragg{0.343};
\def\lbetweenbragg{0.5};

\foreach \x in {0,...,2}
    {\draw[mathematicablue, fill=mathematicablue] (\x*\periodlength,-\gap/2-\pillarradiusy) circle [x radius=\pillarradiusz, y radius=\pillarradiusy];
    \draw[mathematicablue, fill=mathematicablue] (\x*\periodlength,\gap/2+\pillarradiusy) circle [x radius=\pillarradiusz, y radius=\pillarradiusy];}
\draw[mathematicablue, fill=mathematicablue] (-\periodlength/2,\gap/2+2*\pillarradiusy+\braggdistance) rectangle (2.5*\periodlength,\gap/2+2*\pillarradiusy+\braggdistance+\lbragg);
\draw[mathematicablue, fill=mathematicablue] (-\periodlength/2,\gap/2+2*\pillarradiusy+\braggdistance+\lbragg+\lbetweenbragg) rectangle (2.5*\periodlength,\gap/2+2*\pillarradiusy+\braggdistance+2*\lbragg+\lbetweenbragg);
\draw[mathematicablue, fill=mathematicablue] (-\periodlength/2,\gap/2+2*\pillarradiusy+\braggdistance+2*\lbragg+2*\lbetweenbragg) rectangle (2.5*\periodlength,\gap/2+2*\pillarradiusy+\braggdistance+3*\lbragg+2*\lbetweenbragg);
\draw[mathematicablue, fill=mathematicablue] (-\periodlength/2,\gap/2+2*\pillarradiusy+\braggdistance+3*\lbragg+3*\lbetweenbragg) rectangle (2.5*\periodlength,\gap/2+2*\pillarradiusy+\braggdistance+4*\lbragg+3*\lbetweenbragg);
\draw[mathematicablue, fill=mathematicablue] (-\periodlength/2,\gap/2+2*\pillarradiusy+\braggdistance+4*\lbragg+4*\lbetweenbragg) rectangle (2.5*\periodlength,\gap/2+2*\pillarradiusy+\braggdistance+5*\lbragg+4*\lbetweenbragg);

\draw[black, <->] (\periodlength,-\gap/2) -- (\periodlength,\gap/2) node[pos=0.5,right]{\SI{1.2}{\micro\metre}};
\draw[black, <->] (0,-\gap/2-\pillarradiusy) -- (\periodlength,-\gap/2-\pillarradiusy) node[pos=0.5,below]{\SI{2}{\micro\metre}};
\draw[black, <->] (0,\gap/2) -- (0,\gap/2+2*\pillarradiusy) node[pos=0.5,left]{\SI{1.68}{\micro\metre}};
\draw[black, <->] (2*\periodlength-\pillarradiusz,\gap/2+\pillarradiusy) -- (2*\periodlength+\pillarradiusz,\gap/2+\pillarradiusy) node[pos=0.5,above]{\SI{1.75}{\micro\metre}};
\draw[black, <->] (\periodlength,\gap/2+2*\pillarradiusy) -- (\periodlength,\gap/2+2*\pillarradiusy+\braggdistance) node[pos=0.5,left]{\SI{1.84}{\micro\metre}};
\draw[black, <->] (\periodlength,\gap/2+2*\pillarradiusy+\braggdistance+\lbragg+\lbetweenbragg) -- (\periodlength,\gap/2+2*\pillarradiusy+\braggdistance+2*\lbragg+\lbetweenbragg) node[pos=0.5,left]{\SI{342.6}{\nano\metre}};
\draw[black, <->] (\periodlength,\gap/2+2*\pillarradiusy+\braggdistance+3*\lbragg+2*\lbetweenbragg) -- (\periodlength,\gap/2+2*\pillarradiusy+\braggdistance+3*\lbragg+3*\lbetweenbragg) node[pos=0.5,left]{\SI{0.5}{\micro\metre}};

\end{tikzpicture}
\caption{\label{fig:PillarDimensions}Dimensions of the dual pillar structure. The structure is made of fused silica ($\varepsilon_r=2.13$).}

\end{figure}

\begin{figure}[htp]
\begin{tikzpicture}[scale=0.8]
\definecolor{mathematicablue}{rgb}{0.3684,0.5068,0.7098};
\definecolor{mathematicayellow}{rgb}{0.8807,0.611,0.1421};

\def\periodlength{2};
\def\toothwidth{1.2};
\def\toothheight{1.75};
\def\gap{1.2};
\def\basis{2};

\draw[mathematicayellow, fill=mathematicayellow] (-0.5*\periodlength,-\gap/2-\toothheight-\basis) rectangle (2.5*\periodlength,-\gap/2-\toothheight);
\draw[mathematicayellow, fill=mathematicayellow] (-0.5*\periodlength,\gap/2+\toothheight) rectangle (2.5*\periodlength,\gap/2+\toothheight+\basis);
\foreach \x in {0,...,2}
    {\draw[mathematicayellow, fill=mathematicayellow] (\x*\periodlength-\toothwidth/2,-\gap/2-\toothheight) rectangle (\x*\periodlength+\toothwidth/2,-\gap/2);
    \draw[mathematicayellow, fill=mathematicayellow] (\x*\periodlength-\toothwidth/2,\gap/2) rectangle (\x*\periodlength+\toothwidth/2,\gap/2+\toothheight);}

\draw[black, <->] (\periodlength,-\gap/2) -- (\periodlength,\gap/2) node[pos=0.5,right]{\SI{1.2}{\micro\metre}};
\draw[black, <->] (0,-\gap/2-\toothheight-\basis/2) -- (\periodlength,-\gap/2-\toothheight-\basis/2) node[pos=0.5,below]{\SI{2}{\micro\metre}};
\draw[black, <->] (2*\periodlength-\toothwidth/2,-\gap/2-\toothheight/2) -- (2*\periodlength+\toothwidth/2,-\gap/2-\toothheight/2) node[pos=0.5,below]{\SI{1.2}{\micro\metre}};
\draw[black, <->] (0,-\gap/2-\toothheight) -- (0,-\gap/2) node[pos=0.5,left]{\SI{1.75}{\micro\metre}};

\end{tikzpicture}
\caption{\label{fig:RecDimensions}Dimensions of the dual-layer rectangular grating structure. The structure is made of fused silica ($\varepsilon_r=2.13$).}
\end{figure}
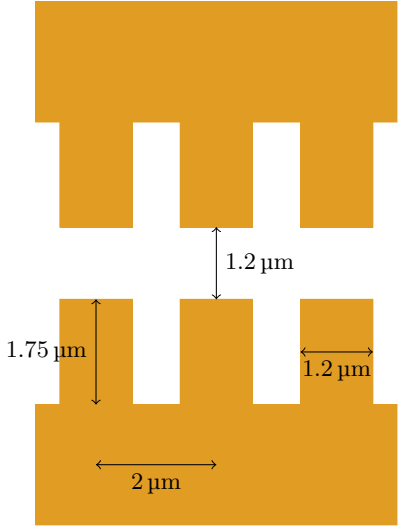

The dimensions of the analyzed DLA structures made of fused silica are optimized for highest gradient at given incoming laser peak field. Fixed dimensions are the vacuum gap and the period length, optimization parameters are the pillar dimensions for the dual pillar structure and the tooth dimensions for the dual-layer rectangular grating structure. If a Bragg mirror is used, the distance between the pillars and the first layer is an additional optimization parameter, the layer thickness and the distance between the layers are, however, given by the relative permittivity of the dielectric. The dimensions of the optimized structures are summarized in Figs~\ref{fig:PillarDimensions} and \ref{fig:RecDimensions}. For wakefield studies with a laser wavelength of \SI{800}{\nano\metre}, the dimensions of the dual pillar structure are scaled by a factor of 0.4. 

\section{\label{sec:app2ParticleModel}Two-particle model}
\subsection{\label{sec:appBeamTransport}Beam transport}
The strong head-tail instability for a bunch passing a circular accelerator is analytically described by Chao's two-particle model~\cite{Chao1993PhysicsAccelerators}. The simplified model in smooth approximation consider a bunch which is made of two macroparticles executing synchrotron oscillations with equal amplitude but opposite phase. During the first half of the synchrotron period, the leading particle has the index 1, the trailing particle the index 2. Chao's derivation assumes that the synchrotron frequency is much smaller than the transverse betatron frequency. According to the APF scheme~\cite{Niedermayer2018Alternating-PhaseAcceleration}, both frequencies are indeed the same and we adapt the model in the following. 

The equations of motion for the two macroparticles in a focusing lattice are
\begin{subequations}
\begin{align}
    y''_1+k_\beta^2 y_1&=0 \label{eq:AppEomFirstParticle}\\
    y''_2+k_\beta^2 y_2&= C_W y_1,\label{eq:AppEomSecondParticle}
\end{align}
\label{eq:AppEom}
\end{subequations}
with
\begin{equation}
    C_W=\frac{q q_\mathrm{bunch}}{p_{z0}\beta_\mathrm{ref}c_0\lambda_0}W'_y,
\end{equation}
the betatron wavenumber $k_\beta$, the particle charge $q$, the bunch charge $q_\mathrm{bunch}$, the reference momentum of the bunch $p_{z0}$, the reference velocity $\beta_\mathrm{ref}$, the speed of light $c_0$, the laser wavelength $\lambda_0$ and the slope $W'_y=\partial W_y / \partial y$ of the transverse wake per grating cell. We assume the wake is a constant longitudinally and linear in the transverse offset. A solution of Eq.~\ref{eq:AppEomFirstParticle} for the leading particle is the unperturbed (betatron) oscillation
\begin{equation}
y_1\left(z\right)=y_1\left(0\right)\cos\left(k_\beta z\right)+\frac{y'_1\left(0\right)}{k_\beta}\sin\left(k_\beta z\right).\label{eq:AppSolFirstParticle}
\end{equation}
The solution of Eq.~\ref{eq:AppEomSecondParticle} for the trailing particle is
\begin{align}
\begin{split}
    y_2\left(z\right)=&\cos\left(k_\beta z\right)\left(y_2\left(0\right)-\frac{y'_1\left(0\right)}{2k_\beta^2}C_Wz\right)\\
    &+\sin\left(k_\beta z\right)\left(\frac{y'_2\left(0\right)}{k_\beta}+\frac{y'_1\left(0\right)}{2k^3_\beta}C_W\right.\\&\left.+\frac{y_1\left(0\right)}{2k_\beta}C_Wz\right).\label{eq:AppSolSecondParticle}
\end{split}
\end{align}
The first terms in the brackets describe the unperturbed (betatron) oscillation of the trailing particle and the second terms, growing linearly with z, describe the perturbation by the wake. The solutions~\ref{eq:AppSolFirstParticle} and~\ref{eq:AppSolSecondParticle} can be combined to a matrix form
\begin{equation}
    \begin{pmatrix}
        y_1\\y'_1\\y_2\\y'_2
    \end{pmatrix}
    =
    \begin{pmatrix}
    M_\beta^f\left(z\right) & 0\\M_W^f\left(z\right) & M_\beta^f\left(z\right)
    \end{pmatrix}
    \begin{pmatrix}
        y_1\\y'_1\\y_2\\y'_2
    \end{pmatrix}_{z=0},\label{eq:AppMatrix}
\end{equation}
where the entries in the matrix are themselves 2x2 matrices given by
\begin{equation}
    M_\beta^f\left(z\right)=
    \begin{pmatrix}
    \cos{\left(k_\beta z\right)} & \frac{1}{k_\beta}\sin{\left(k_\beta z\right)}\\-k_\beta\sin{\left(k_\beta z\right)} &  \cos{\left(k_\beta z\right)}
    \end{pmatrix}\label{eq:AppMatrixFbeta}
\end{equation}
and
\begin{widetext}
\begin{equation}
    M_W^f\left(z\right)=
    \begin{pmatrix}
    \frac{C_W}{2k_\beta}z\sin{\left(k_\beta z\right)} & \frac{C_W}{2k_\beta^3}\sin{\left(k_\beta z\right)}-\frac{C_W}{2k_\beta^2}z\cos{\left(k_\beta z\right)}\\\frac{C_W}{2k_\beta}\sin{\left(k_\beta z\right)}+\frac{C_W}{2}z\cos{\left(k_\beta z\right)}  &  \frac{C_W}{2k_\beta}z\sin{\left(k_\beta z\right)}
    \end{pmatrix}.
    \label{eq:AppMatrixFW}
\end{equation}
\end{widetext}
Evaluating Eqs~\ref{eq:AppMatrixFbeta} and~\ref{eq:AppMatrixFW} after half a longitudinal oscillation period and taking into account that the longitudinal oscillation period equals the transverse betatron period, the transfer matrices become
\begin{equation}
    M_\beta^f\left(\frac{L_\beta}{2}\right)=
    \begin{pmatrix}
    -1 & 0\\0 &  -1
    \end{pmatrix}
\end{equation}
and
\begin{equation}
    M_W^f\left(\frac{L_\beta}{2}\right)=
    \begin{pmatrix}
   0 & \frac{C_W}{16\pi^2}L_\beta^3\\-\frac{C_W}{4}L_\beta  &  0
    \end{pmatrix}.
\end{equation}
The results for the second half oscillation period where the particles change there position can be obtained by exchanging the indices in Eq.~\ref{eq:AppMatrix}. The transfer matrix of a full oscillation period is then given by
\begin{equation}
    \begin{pmatrix}
        y_1\\y'_1\\y_2\\y'_2
    \end{pmatrix}_{z=L_\beta}
    =
    \begin{pmatrix}
    1+M_W^f\left(\frac{L_\beta}{2}\right)^2 & M_W^f\left(\frac{L_\beta}{2}\right)\\M_W^f\left(\frac{L_\beta}{2}\right) & 1
    \end{pmatrix}
    \begin{pmatrix}
        y_1\\y'_1\\y_2\\y'_2
    \end{pmatrix}_{z=0}.
\end{equation}
We can now eigenanalyze the resulting transfer matrix according to Chao's stability analysis. Stability requires that for all solutions of the eigenvalues $\lambda$ the function $\lambda+1/\lambda$ is real and its value is between -2 and 2. This gives the stability criterion
\begin{equation}
    C_W\leq\frac{16\pi}{L_\beta^2},
    \label{eq:AppStabilitySmooth}
\end{equation}
which can be used as an estimation for the maximum bunch charge in a APF grating (cf. Sec.~\ref{sec:stabilization} and Fig.~\ref{fig:yOffsetStabilitySmooth}).

Calculating the full APF transfer matrices without the smooth approximation requires also knowledge of the transfer matrices for the defocusing sections in an APF grating. The equations of motion in these sections are equal to the Eqs.~\ref{eq:AppEom} except that the sign changes. This results in exchanging the trigonometric functions in Eqs.~\ref{eq:AppMatrixFbeta} and~\ref{eq:AppMatrixFW} by hyperbolic functions which yields
\begin{equation}
    M_\beta^d\left(z\right)=
    \begin{pmatrix}
    \cosh{\left(k_\beta z\right)} & \frac{1}{k_\beta}\sinh{\left(k_\beta z\right)}\\k_\beta\sinh{\left(k_\beta z\right)} &  \cosh{\left(k_\beta z\right)}
    \end{pmatrix}
\end{equation}
and
\begin{widetext}
\begin{equation}
    M_W^d\left(z\right)=
    \begin{pmatrix}
    \frac{C_W}{2k_\beta}z\sinh{\left(k_\beta z\right)} & \frac{C_W}{2k_\beta^2}z\cosh{\left(k_\beta z\right)}-\frac{C_W}{2k_\beta^3}\sinh{\left(k_\beta z\right)}\\
    \frac{C_W}{2k_\beta}\sinh{\left(k_\beta z\right)}+\frac{C_W}{2}z\cosh{\left(k_\beta z\right)}  &  \frac{C_W}{2k_\beta}z\sinh{\left(k_\beta z\right)}
    \end{pmatrix}.
    \label{eq:AppMatrixDW}
\end{equation}
\end{widetext}
Tracking with these transfer matrices describes the stability behavior significantly better than with the smooth approximation (cf. Sec.~\ref{sec:stabilization} and Fig.~\ref{fig:yOffsetStabilityAPF}). Calculating an analytical stability threshold as in Eq.~\ref{eq:AppStabilitySmooth} is, however, not possible.

\subsection{Acceleration}
\label{sec:app2}
Considering coherent acceleration of a bunch the bunch energy is a function of time, thus equivalently of the longitudinal position in the grating. Using again a two-particle model, the transverse equations of motion for the two macroparticles are given by
\begin{subequations}
\begin{align}
    \frac{d}{dz}\left(\gamma\left(z\right)\frac{dy_1}{dz}\right)+\gamma\left(z\right)k_\beta\left(z\right)^2 y_1&=0\\
    \frac{d}{dz}\left(\gamma\left(z\right)\frac{dy_2}{dz}\right)+\gamma\left(z\right)k_\beta\left(z\right)^2 y_2&= \gamma\left(z\right) C_W\left(z\right) y_1,
\end{align}
\end{subequations}
where the Lorentz factor $\gamma\left(z\right)$ and thus also the wake term on the right hand side are functions of the longitudinal position $z$. For a constant phase advance per FD-cell, the FD-cell length has to be proportional to $\gamma^{3/2}$ according to the energy dependence of the focusing strength. This means
\begin{equation}
    k_\beta\left(z\right)=\frac{k_{\beta,0}}{\gamma\left(z\right)^{3/2}}\gamma_0^{3/2}.
\end{equation}
If we consider a linear energy gain ${\gamma\left(z\right)=\gamma_0\left(1+G\cos{\phi_s}/\left(m_ec_0^2\right) z\right)}$ and apply the transformation $u=1+\alpha z$ with $\alpha=G\cos{\phi_s}/\left(m_ec_0^2\right)$, the equations of motion become
\begin{subequations}
\begin{align}
    \frac{d^2y_1}{du^2}+\frac{1}{u}\frac{dy_1}{du}+\frac{k_\gamma^2}{u^3} y_1&=0\label{eq:AppAccy1}\\
   \frac{d^2y_2}{du^2}+\frac{1}{u}\frac{dy_2}{du}+\frac{k_\gamma^2}{u^3} y_2&= \frac{C_{W,0}}{\alpha^2 u} y_1.\label{eq:AppAccy2}
\end{align}
\end{subequations}
with $k_\gamma=k_{\beta,0}/\alpha$. A solution of Eq.~\ref{eq:AppAccy1} is
\begin{equation}
    y_1\left(u\right)=c_1J_0\left(\frac{2k_\gamma}{\sqrt{u}}\right)+c_2N_0\left(\frac{2k_\gamma}{\sqrt{u}}\right),    
\end{equation}
where $J_0\left(x\right)$ and $N_0\left(x\right)$ are Bessel and Neumann functions. Using the asymptotic expressions and the initial conditions $y_1\left(u=1\right)=y_0$ and $y'_1\left(u=1\right)=y'_0$, the solution of the first macroparticle becomes
\begin{align}
\begin{split}
    y_1\left(u\right)=&\sqrt[4]{u}\left[y_0\cos\left(2k_\gamma\left(1-\frac{1}{\sqrt{u}}\right)\right)\right.\\&-\left.\frac{\alpha y_0-4y'_0}{4\alpha k_\gamma}\sin\left(2k_\gamma\left(1-\frac{1}{\sqrt{u}}\right)\right)\right].
    \label{eq:AppSolutiony1}
\end{split}
\end{align}
Equation~\ref{eq:AppSolutiony1} shows that the betatron oscillations in a DLA grating are not damped adiabatically compared to a conventional magnetic focusing lattice. An increase of the stability threshold as described by Chao for the conventional magnetic focusing lattice can thus not be expected and numerical tracking simulations confirm this statement (cf.~\ref{sec:acceleration}). For the sake of completeness, Eq.~\ref{eq:AppAccy2} can be solved as
\begin{equation}
    y_2\left(u\right)=y_1\left(u\right)+\frac{C_{W,0}}{\alpha^2}\int_1^u G\left(u,\tilde{u}\right)y_1\left(\tilde{u}\right)d\tilde{u},
    \label{eq:AppIntegral}
\end{equation}
where the Green's function is given by
\begin{subequations}
\begin{align}
    \begin{split}
    G\left(u,\tilde{u}\right)=&-\pi\left[J_0\left(\frac{2k_\gamma}{\sqrt{u}}\right)N_0\left(\frac{2k_\gamma}{\sqrt{\tilde{u}}}\right)\right.\\
    &\left.-N_0\left(\frac{2k_\gamma}{\sqrt{u}}\right)J_0\left(\frac{2k_\gamma}{\sqrt{\tilde{u}}}\right)\right]
    \end{split}\\
    \approx&\frac{\sqrt[4]{u\tilde{u}}}{k_\gamma}\sin\left(2k_\gamma\left(\frac{1}{\sqrt{u}}-\frac{1}{\sqrt{\tilde{u}}}\right)\right).
\end{align}
\end{subequations}

\bibliography{bibliography}

\end{document}